\documentclass{article}
\usepackage{spconf,amsmath,graphicx}

\usepackage{enumitem}
\setlist{nosep, leftmargin=14pt}
\usepackage[colorlinks=true, allcolors=blue]{hyperref}
\usepackage{mwe} 

\def\r{{\mathbf r}}

\title{Towards an accurate and generalizable multiple sclerosis lesion segmentation model using self-ensembled lesion fusion}
\address{$^{\star}$Johns Hopkins University,
    $^{\ddagger}$National~Institutes~of~Health}

\begin{document}
%
\maketitle
\begin{abstract}
Automatic multiple sclerosis~(MS) lesion segmentation using multi-contrast magnetic resonance~(MR) images provides improved efficiency and reproducibility compared to manual delineation.
Current state-of-the-art automatic MS lesion segmentation methods utilize modified U-Net-like architectures.
However, in the literature, dedicated architecture modifications were always required to maximize their performance.
In addition, the best-performing methods have not proven to be generalizable to diverse test datasets with contrast variations and image artifacts.
In this work, we developed an accurate and generalizable MS lesion segmentation model using the well-known U-Net architecture without further modification.
A novel test-time self-ensembled lesion fusion strategy is proposed that not only achieved the best performance using the ISBI 2015 MS segmentation challenge data but also demonstrated robustness across various self-ensemble parameter choices.
Moreover, equipped with instance normalization rather than batch normalization widely used in literature, the model trained on the ISBI challenge data generalized well on clinical test datasets from different scanners.

\end{abstract}
\begin{keywords}
Self-Ensemble, Domain Generalization, Multiple Sclerosis Lesion, Image Segmentation
\end{keywords}
\section{Introduction}
\label{sec:intro}

Multiple sclerosis~(MS) is characterized by chronic inflammatory demyelination and axonal and neuronal degeneration of the central nervous system~\cite{haider2016topograpy}. 
White matter MS lesions typically appear hyperintense on T2-w and T2-FLAIR MR images, representing focal inflammation in the brain. 
The manual delineation of MS lesions is a time-consuming and subjective process for a trained radiologist, primarily due to the heterogeneous shape, size, and location of these lesions. 
Therefore, developing automatic lesion segmentation methods has been an active research topic for many years.

\begin{figure}[!tb]

\begin{minipage}[b]{1.0\linewidth}
  \centering
  \centerline{\includegraphics[width=8.5cm]{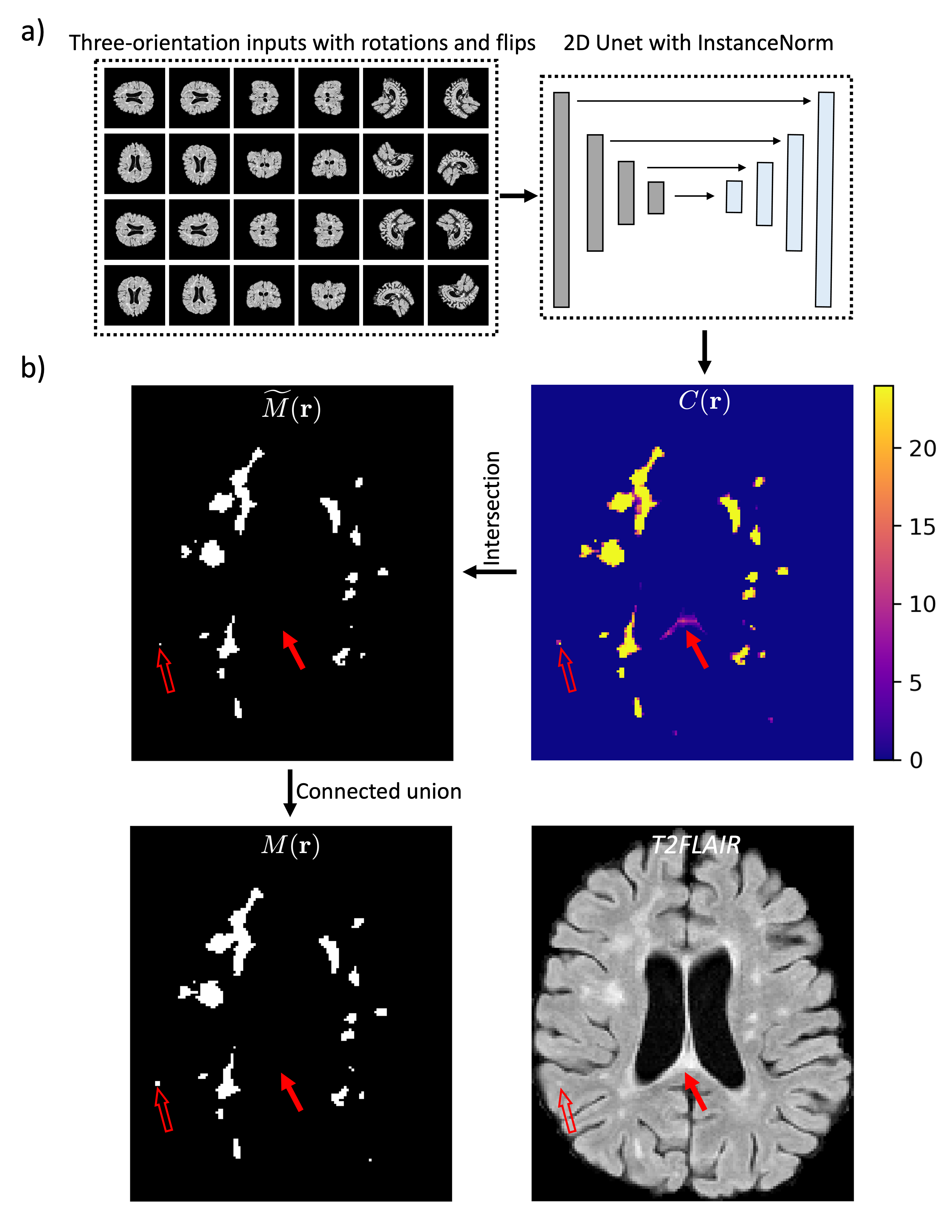}}
\end{minipage}
\caption{Illustration of the proposed self-ensembled lesion fusion (SELF) strategy.}
\label{fig:res1}
\end{figure}

Among all MS lesion segmentation methods, deep learning based methods have been shown to achieve state-of-the-art (SOTA) performance~\cite{zhang2019multiple, zhang2021all}.
These methods typically adopt a widely used U-Net~\cite{ronneberger2015u} architecture with the advantages of multi-scale feature representation and skip connection.
For instance, the best-performing published lesion segmentation methods evaluated using the ISBI longitudinal MS lesion segmentation challenge data~\cite{carass2017longitudinal, carass2017dib} are Tiramisu~\cite{zhang2019multiple} and ALL-Net~\cite{zhang2021all}.
In Tiramisu, the convolutional layer in U-Net was replaced with a densely connected convolutional layer. 
In ALL-Net, a coordinate convolutional layer was incorporated into U-Net.
Tiramisu and ALL-Net are the only two published methods with a score above 93.3 in the ISBI challenge, demonstrating their success in the network architecture designs.

In addition to achieving high accuracy on in-domain tests like the ISBI challenge, addressing out-of-domain shifts during testing is equally important for the reliable clinical deployment of the trained segmentation model.
In clinical settings, one of the challenges is achieving consistent segmentation performance across different MR scanners and imaging protocols.
Several methods have been proposed to improve the cross-domain performance of trained networks, such as cross-scanner data harmonization~\cite{dewey2019deepharmony, zuo2023haca3, hays2023exploring, gebre2023cross}, harmonization-enriched domain adaptation~\cite{zhang2023harmonization}, generic and locally specialized networks~\cite{kamraoui2022deeplesionbrain}, and scanner invariant feature learning~\cite{aslani2020scanner}, contributing to more reliable automatic MS lesion segmentation in clinical settings.

In this paper, we propose a simple yet effective method to achieve both in-domain accuracy and out-of-domain generalization for MS lesion segmentation:
\begin{itemize}
    \item We show that a simple U-Net as the baseline method already achieves good segmentation performance.
    \item We demonstrate that instance normalization generalizes better than the widely used batch normalization in the MS lesion segmentation literature.
    \item Based on the trained U-Net, we deploy a novel test-time Self-Ensembled Lesion Fusion (SELF) strategy to get both SOTA performance in the ISBI challenge and good generalization on several out-of-domain datasets.
\end{itemize}

\begin{figure}[!tb]

\begin{minipage}[b]{1.0\linewidth}
  \centering
  \centerline{\includegraphics[width=8.5cm]{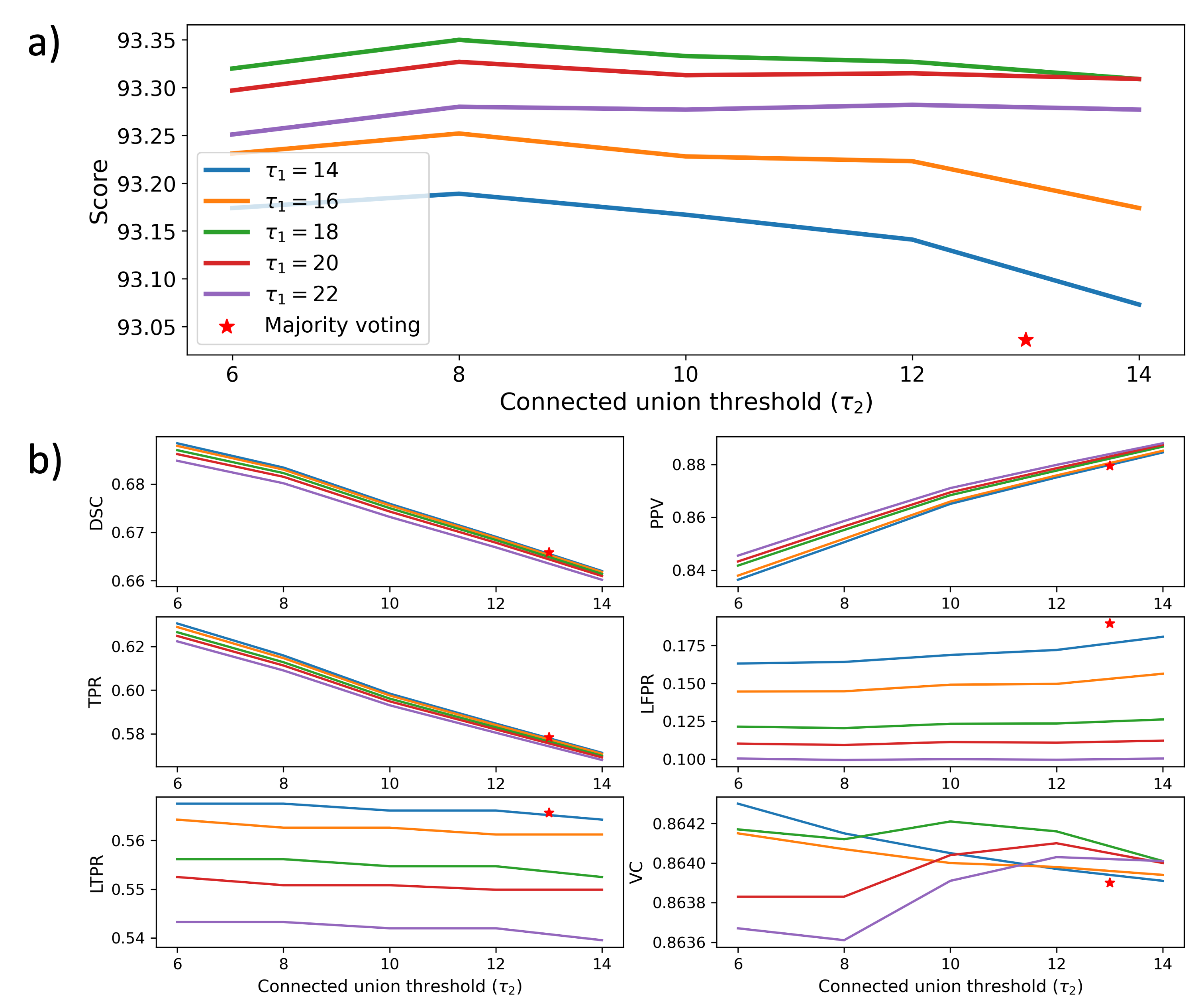}}
\end{minipage}
\caption{The robustness of SELF $\tau_1$ and $\tau_2$ values on the averaged score (a) and their impact on individual metrics (b) in the ISBI challenge. For comparison, we also present majority voting ($\tau_1=\tau_2=13$, marked with red stars).}
\label{fig:res2}
\end{figure}

\section{Methodology}
\label{sec:method}

\subsection{Training Phase}
\label{ssec:training_strategy}

The model takes, as inputs, a stack of three adjacent multi-contrast 2D slices from either the axial, sagittal, or coronal planes.
The output is the predicted lesion mask for the corresponding central slice~\cite{zhang2019multiple}.

There are three key differences between the proposed method and those from past SOTA lesion segmentation methods.
The first difference is the utilization of a simple 2D U-Net architecture without further modification. 
In contrast, ALL-Net and Tiramisu apply coordinate convolution or densely connected convolution layers.
The simplicity of the 2D U-Net enables straightforward implementation with an off-the-shelf segmentation model without a dedicated architecture design.
The second difference is the adoption of instance normalization~\cite{ulyanov2016instance} in the model instead of the batch normalization~\cite{ioffe2015batch} used in ALL-Net and Tiramisu. 
Instance normalization removes instance-specific contrast information from the content image~\cite{ulyanov2016instance}, improving generalization to MR datasets with contrast variations among scanners and protocols ~\cite{han2020automatic}.
The third difference is related to the proposed SELF strategy described in the next section. 
During training, stacks of 2D slices with random orientations (axial, sagittal, or coronal) are extracted from the 3D MR image volumes.
Then random rotations of $0^{\circ}, 90^{\circ}, 180^{\circ} \text{ or } 270^{\circ}$, as well as random flips along either the vertical or horizontal axis, are applied to the extracted 2D slices (Fig.~\ref{fig:res1}a).
Unlike our approach, Tiramisu and ALL-Net use only three orientations for training, without incorporating random rotations or flips.

\subsection{Self-Ensembled Lesion Fusion (SELF)}
During inference, the same 4 rotations and 2 flips are applied to the 3-orientation stacked 2D slice inputs, generating $N_M = 4\times2\times3=24$ binary segmentation masks $M_i(\r)$, where $1 \leq i \leq N_M$ is the index of masks flipped and rotated back into the original space and $\r$ is the spatial coordinate.
A confidence map $C(\r)$ with integer values between $0$ and $N_M$ (Fig.~\ref{fig:res1}b) is generated by adding $N_M$ segmentation masks: 
\begin{equation}
C(\r) = \sum_{i=1}^{N_M} M_i(\r), \ \forall \r.
\end{equation}
For example, if a voxel $\r$ yields $C(\r)=24$, then all rotations and flips of the data agree that $\r$ corresponds to a lesion.
A two-step SELF strategy is then applied to $C(\r)$, involving `intersection' followed by `connected union' as described next.

\subsubsection{Intersection}
\label{sssec:intersection}

The intersection step applies a `voter' threshold $\tau_1$ on $C(\r)$ to generate an initial fused segmentation $\widetilde{M}(\r)$ : 
\begin{equation}
\label{intersec}
\widetilde{M}(\r) = C(\r) > \tau_1, \ \forall \r.
\end{equation}
Fig.~\ref{fig:res1}b shows the resulting $\widetilde{M}(\r)$ by Eq.~\eqref{intersec}, where unconfident segmentations, supported only by a limited number of voters (highlighted by red solid arrows in Fig.~\ref{fig:res1}b), are filtered out after intersection thresholding, thereby preventing over-segmentation of lesions and reducing false positives.

\subsubsection{Connected Union}
\label{sssec:union}
The connected union step first generates a second fused segmentation 
$\widehat{M}(\r)$ from $C(\r)$ with a lower threshold $\tau_2 < \tau_1$:
\begin{equation}
\label{intersec2}
\widehat{M}(\r) = C(\r) > \tau_2, \ \forall \r.
\end{equation}
Then, the final segmentation $M(\r)$ is generated by adding additional segmented locations from $\widehat{M}(\r)$ into $\widetilde{M}(\r)$ if the former are spatially connected to the latter:
\begin{equation}
\label{union}
M(\r) = 
\begin{cases} 
\widehat{M}(\r)& \text{if } \exists \r', \widetilde{M}(\r') >0, \text{s.t. } d_{\widehat{M}}(\r, \r') < \infty, \\
\widetilde{M}(\r) & \text{otherwise},
\end{cases}
\end{equation}
where $d_{\widehat{M}}(\r, \r')$ denotes the distance between locations $\r$ and $\r'$ in $\widehat{M}$. 
This distance is smaller than infinity if and only if $\r$ and $\r'$ belong to the same connected component (i.e., same predicted lesion) in $\widehat{M}$.
Fig.~\ref{fig:res1}b shows the final segmentation $M(\r)$ by Eq.~\eqref{union}, where delineation of the small lesion (highlighted by red hollow arrows in Fig.~\ref{fig:res1}b) is enlarged and further refined from `Intersection', improving segmentation accuracy.
A similar double thresholding idea has been proposed and proven to be successful in Canny edge detection~\cite{canny1986computational}.
We will show its effectiveness on MS lesion segmentation below.

\begin{figure}[!tb]

\begin{minipage}[b]{1.0\linewidth}
  \centering
  \centerline{\includegraphics[width=8.5cm]{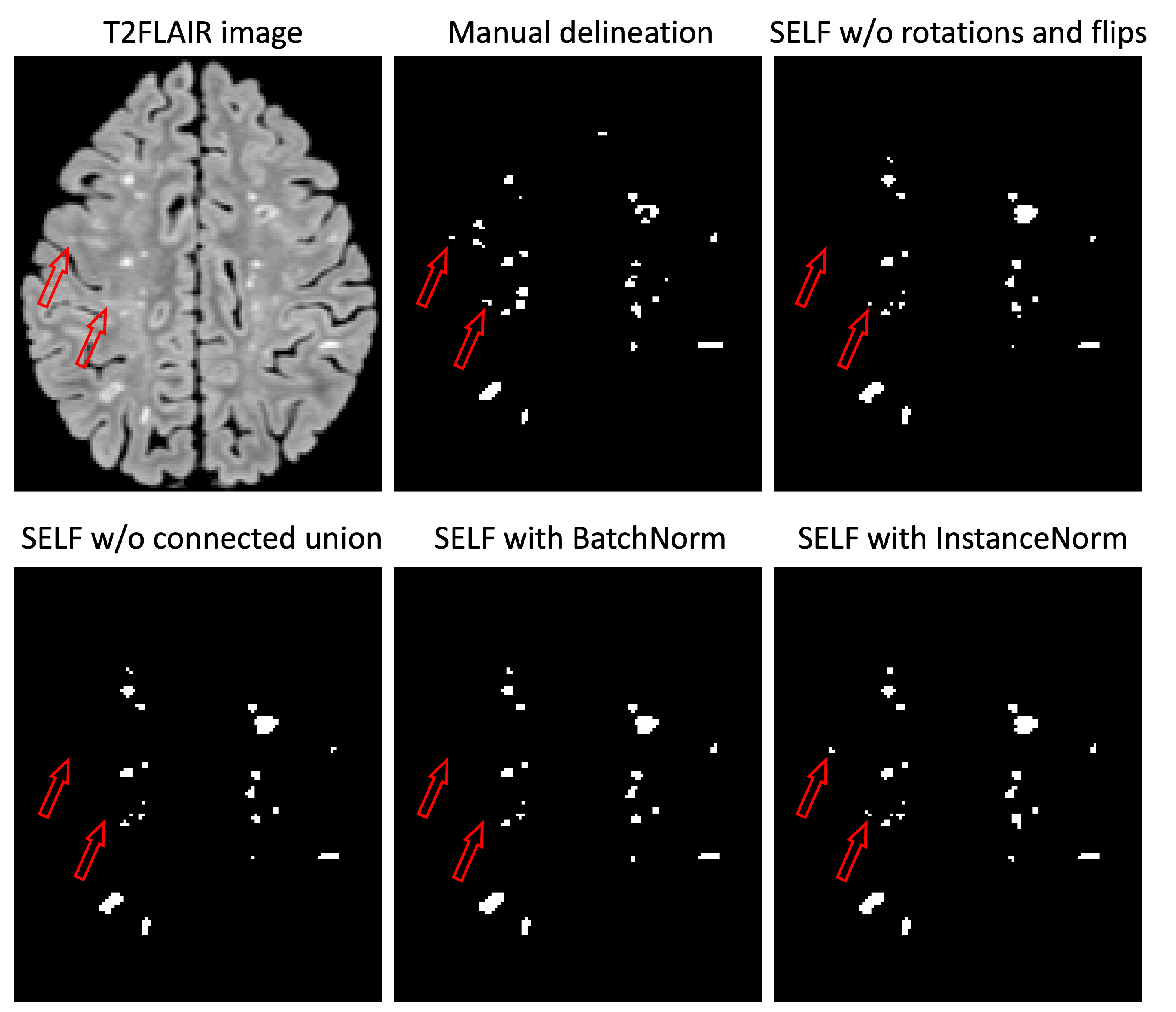}}
\end{minipage}
\caption{Segmentation results of ISBI trained model on one out-of-domain test subject with manual delineation reference.}
\label{fig:res3}
\end{figure}

\section{Experimental Results}
\label{sec:results}
All models were trained on the ISBI challenge training dataset and tested on ISBI and multiple out-of-domain test datasets.
For the ISBI data, 21 and 61 longitudinal multi-contrast MR images from 5 and 14 patients were included in the training and test datasets, respectively.
All the ISBI data were rigidly registered to MNI space using MIPAV’s optimized automatic registration\footnote{\url{https://mipav.cit.nih.gov/}} followed by brain and dura extraction\footnote{\url{https://www.nitrc.org/projects/toads-cruise/}} and N4 bias field correction\footnote{\url{https://stnava.github.io/ANTs/}}.
All the out-of-domain test data went through similar pre-processing steps as the ISBI data with an additional super-resolution step using SMORE~\cite{remedios2023self, zhao2020smore} before registration to MNI space.

We used $L_2$ loss between the predicted probabilistic segmentation map and the label for back-propagation~\cite{zhang2019multiple}.
Affine and elastic augmentations were applied to multi-contrast 3D volumes during training.
The Adam optimizer~\cite{kingma2014adam} with an initial learning rate $10^{-4}$, mini-batch size $12$, and $35000$ training iterations was used to update the network's weights.

Evaluation metrics were based on those used in the ISBI 2015 challenge~\cite{carass2017longitudinal}: Dice Similarity Coefficient~(DSC), Precision~(PPV), Sensitivity~(TPR), Lesion-wise True Positive Rate~(LTPR), Lesion-wise False Positive Rate~(LFPR) and Pearson’s correlation coefficient of the lesion volumes~(VC).
A challenge score based on the weighted average of the above metrics~\cite{carass2017longitudinal} was also presented on ISBI test results.

\begin{table}[!tb]
\centering
\caption{Performance Comparison on ISBI 2015 Challenge. The best and second-best performances are indicated in \textbf{bold} and \underline{underline}, respectively.}
\setlength{\tabcolsep}{2.5pt} 
\small 
\begin{tabular}{cccccccc}
\hline
Algorithms & Score & DSC & PPV & TPR & LFPR & LTPR & VC\\
\hline
SELF (proposed) & \bf{93.35} & \bf{0.682} & 0.855 & \bf{0.613} & \bf{0.121} & \bf{0.542} & \underline{0.864} \\
All-Net~\cite{zhang2021all} & \underline{93.32} & 0.639 & \bf{0.914} & 0.525 & \underline{0.122} & \underline{0.533} & 0.860 \\
Tiramisu~\cite{zhang2019multiple} & 93.21 & \underline{0.643} & \underline{0.908} & \underline{0.533} & 0.124 & 0.520 & \bf{0.867} \\
\hline
\end{tabular}
\label{tab:res1}
\end{table}

\begin{table}[!tb]
\centering
\caption{Ablation Study on ISBI 2015 Challenge.}
\setlength{\tabcolsep}{2pt} 
\small 
\begin{tabular}{cccccccc}
\hline
Algorithms & Score & DSC & PPV & TPR & LFPR & LTPR & VC\\
\hline
w/o rotations and flips & 92.83 & \underline{0.668} & 0.872 & \underline{0.583} & 0.224 & \bf{0.575} & \underline{0.864} \\
w/o connected union & 93.18 & 0.639 & \bf{0.906} & 0.533 & 0.137 & 0.533 & 0.862 \\
w/ BatchNorm & \underline{93.25} & 0.655 & \underline{0.886} & 0.565 & \bf{0.109} & 0.516 & 0.861 \\
w/ InstanceNorm & \bf{93.35} & \bf{0.682} & 0.855 & \bf{0.613} & \underline{0.121} & \underline{0.542} & \bf{0.864} \\
 
\hline
\end{tabular}
\label{tab:res2}
\end{table}

\begin{table}[!tb]
\centering
\caption{Performance on Out-of-domain Test Dataset.}
\setlength{\tabcolsep}{2.5pt} 
\small 
\begin{tabular}{cccccccc}
\hline
Algorithms & DSC & PPV & TPR & LFPR & LTPR & VC\\
\hline
w/o rotations and flips & 0.614 & \underline{0.809} & 0.523 & 0.154 & \bf{0.375} & 0.978 \\
w/o connected union & 0.605 & \bf{0.823} & 0.507 & \underline{0.124} & 0.361 & \underline{0.979} \\
w/ BatchNorm & \underline{0.618} & 0.803 & \underline{0.529} & 0.127 & 0.321 & \bf{0.980} \\
w/ InstanceNorm & \bf{0.633} & 0.759 & \bf{0.576} & \bf{0.104} & \underline{0.368} & 0.974 \\
 
\hline
\end{tabular}
\label{tab:res3}
\end{table}

\begin{figure*}[!tb]
    \centering
    \includegraphics[width=\linewidth]{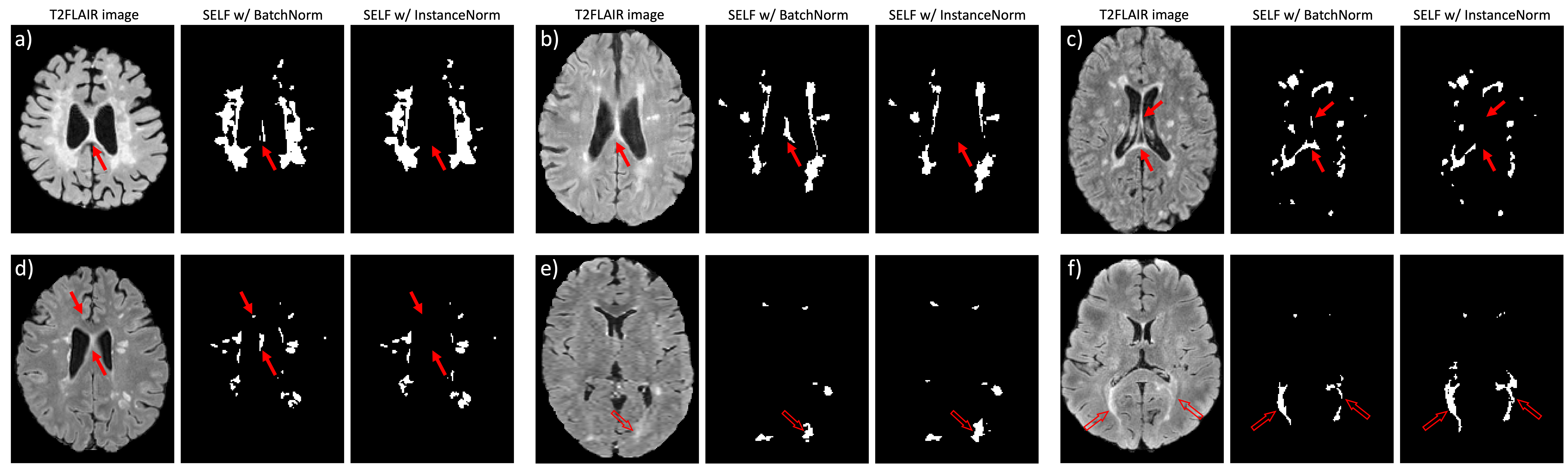}
    \caption{Segmentation results of ISBI trained model on six test subjects under contrast variations and imaging artifacts.}
    \label{fig:res4}
\end{figure*}

\subsection{Robustness of SELF on ISBI Challenge}
Fig.~\ref{fig:res2} illustrates the robustness of SELF $\tau_1$ (Eq.~\eqref{intersec}) and $\tau_2$ (Eq.~\eqref{intersec2}) values on the averaged score (Fig.~\ref{fig:res2}a) in the ISBI challenge and their impact on individual metrics (Fig.~\ref{fig:res2}b).
In Fig.~\ref{fig:res2}a, the optimal score of $93.35$ was achieved with $\tau_1=18 \text{ and } \tau_2=8$.
Consistent scores above $93.25$ were achieved for $18 \leq \tau_1 \leq 22$ and $6 \leq \tau_2 \leq 14$, demonstrating the robustness of SELF on the ISBI challenge test dataset.
In Fig.~\ref{fig:res2}b, changes in $\tau_1$ at each fixed $\tau_2$ had relatively little impact on DSC, PPV, and TPR.
Larger $\tau_2$ contributed to improved PPV, while at the expense of decreased DSC and TPR.
A possible explanation could be that since DSC, PPV and TPR are voxel-wise metrics, they are more susceptible to the total volume of lesions segmented, which is mainly affected by $\tau_2$.
Conversely, LFPR and LTPR were mainly affected by $\tau_1$. Larger $\tau_1$ led to improved LFPR but worse LTPR.
This is because LFPR and LTPR are lesion-wise metrics determined by the number of individually segmented lesions, which is mainly affected by $\tau_1$.
VC is invariant of $\tau_1$ and $\tau_2$.

\subsection{Comparison and Ablation on ISBI Challenge}
We compared SELF to ALL-Net~\cite{zhang2021all} and Tiramisu~\cite{zhang2019multiple} with metrics reported in their papers.
Performance metrics are shown in Table~\ref{tab:res1}.
SELF performed similarly or even better than ALL-Net and Tiramisu on many metrics.
A noticeably lower PPV, higher TPR, and DSC were observed in SELF compared to ALL-Net and Tiramisu.
This suggests that SELF with $\tau_1=18 \text{ and } \tau_2=8$, despite achieving the highest score, tends to be more `aggressive' in capturing lesions.

An ablation study regarding SELF and instance normalization was also shown in Table~\ref{tab:res2}.
For SELF, improved performance was observed by adding rotations and flips and the second step `connected union'.
For instance normalization, slightly worse performance was observed when switching to batch normalization as used in ALL-Net and Tiramisu.

\subsection{Out-of-domain Test Dataset with Labels}
We tested the generalization of SELF trained on ISBI using one out-of-domain dataset, which consists of 10 patients with manual delineation of MS lesions by an experienced radiologist.
Similar to the ISBI ablation study, we compared to three orientation inputs without rotations and flips (similar to Tiramisu), intersection without connected union, and batch normalization instead of instance normalization.
The source code of ALL-Net is not made public and therefore not compared here.
Table~\ref{tab:res3} shows the performance metrics, presenting consistent improvement and accurate segmentation as in Table~\ref{tab:res2}. 
Fig.~\ref{fig:res3} shows a representative test subject with multiple small lesions.
With manual delineation as a reference, more lesions (highlighted by red hollow arrows) were captured by SELF with instance normalization.

\subsection{More Out-of-domain Test Datasets}
We further tested the generalization of SELF trained on ISBI using more out-of-domain test data acquired from different scanners. 
Fig.~\ref{fig:res4} shows the generalization performance of instance and batch normalizations of SELF under contrast variation (Figs.~\ref{fig:res4}a-f), motion artifacts (Figs.~\ref{fig:res4}b and ~\ref{fig:res4}c), and noise (Fig.~\ref{fig:res4}e) among scanners.
In Figs.~\ref{fig:res4}a-d, false positives (red solid arrows) were shown in batch normalization, which were not shown in instance normalization.
In Figs.~\ref{fig:res4}e-f, under-segmentations of lesions (red hollow arrows) were observed on batch normalization but were corrected on instance normalization.
Qualitatively, SELF with instance normalization achieved accurate and robust segmentations under various acquisition conditions in clinics. 

\section{Conclusion}
\label{sec:conclusion}
We propose SELF, a test-time self-ensembled lesion fusion strategy with a simple U-Net architecture using instance normalization to achieve both accurate and generalizable segmentation performance for MS lesions using multi-contrast MR images.
Future work includes a systematic validation of the generalization ability of SELF with large-scale clinical datasets.

\section{Compliance with ethical standards}
\label{sec:ethics}

This research study was conducted retrospectively using human subject data following an IRB approved protocol.

\section{Acknowledgments}
\label{sec:acknowledgments}

This material is partially supported by the National Science Foundation Graduate Research Fellowship under Grant No. DGE-1746891 (Remedios).
This work also received support from National Multiple Sclerosis Society RG-1907-34570 (Pham) and CDMRP W81XWH2010912 (Prince).
The opinions and assertions expressed herein are those of the authors and do not reflect the official policy or position of the Uniformed Services University of the Health Sciences or the Department of Defense.

\bibliographystyle{IEEEbib}
\bibliography{strings,refs}

\begin{thebibliography}{10}

\bibitem{haider2016topograpy}
L.~Haider et~al.,
\newblock ``The topograpy of demyelination and neurodegeneration in the multiple sclerosis brain,''
\newblock {\em Brain}, vol. 139, no. 3, pp. 807--815, 2016.

\bibitem{zhang2019multiple}
H.~Zhang et~al.,
\newblock ``{Multiple sclerosis lesion segmentation with tiramisu and 2.5D stacked slices},''
\newblock in {\em Medical Image Computing and Computer Assisted Intervention--MICCAI 2019: 22nd International Conference, Shenzhen, China, October 13--17, 2019, Proceedings, Part III 22}. Springer, 2019, pp. 338--346.

\bibitem{zhang2021all}
H.~Zhang et~al.,
\newblock ``{ALL-Net: Anatomical information lesion-wise loss function integrated into neural network for multiple sclerosis lesion segmentation},''
\newblock {\em NeuroImage: Clinical}, vol. 32, pp. 102854, 2021.

\bibitem{ronneberger2015u}
O.~Ronneberger et~al.,
\newblock ``{U-net: Convolutional networks for biomedical image segmentation},''
\newblock in {\em Medical Image Computing and Computer-Assisted Intervention--MICCAI 2015: 18th International Conference, Munich, Germany, October 5-9, 2015, Proceedings, Part III 18}. Springer, 2015, pp. 234--241.

\bibitem{carass2017longitudinal}
A.~Carass et~al.,
\newblock ``{Longitudinal multiple sclerosis lesion segmentation: Resource and challenge},''
\newblock {\em NeuroImage}, vol. 148, pp. 77--102, 2017.

\bibitem{carass2017dib}
A.~Carass et~al.,
\newblock ``{Longitudinal multiple sclerosis lesion segmentation data resource},''
\newblock vol. 12, pp. 346--350, 2017.

\bibitem{dewey2019deepharmony}
B.E. Dewey et~al.,
\newblock ``{DeepHarmony: A deep learning approach to contrast harmonization across scanner changes},''
\newblock {\em Magnetic Resonance Imaging}, vol. 64, pp. 160--170, 2019.

\bibitem{zuo2023haca3}
L.~Zuo et~al.,
\newblock ``{HACA3: A unified approach for multi-site MR image harmonization},''
\newblock {\em Computerized Medical Imaging and Graphics}, vol. 109, pp. 102285, 2023.

\bibitem{hays2023exploring}
S.~Hays et~al.,
\newblock ``Exploring the optimal operating mr contrast for brain ventricle parcellation,''
\newblock in {\em Medical Imaging with Deep Learning, short paper track}, 2023.

\bibitem{gebre2023cross}
Robel~K Gebre et~al.,
\newblock ``Cross--scanner harmonization methods for structural mri may need further work: A comparison study,''
\newblock {\em Neuroimage}, vol. 269, pp. 119912, 2023.

\bibitem{zhang2023harmonization}
J.~Zhang et~al.,
\newblock ``{Harmonization-enriched domain adaptation with light fine-tuning for multiple sclerosis lesion segmentation},''
\newblock {\em arXiv preprint arXiv:2310.20586}, 2023.

\bibitem{kamraoui2022deeplesionbrain}
R.A. Kamraoui et~al.,
\newblock ``{DeepLesionBrain: Towards a broader deep-learning generalization for multiple sclerosis lesion segmentation},''
\newblock {\em Medical Image Analysis}, vol. 76, pp. 102312, 2022.

\bibitem{aslani2020scanner}
S.~Aslani et~al.,
\newblock ``{Scanner invariant multiple sclerosis lesion segmentation from MRI},''
\newblock in {\em 2020 IEEE 17th International Symposium on Biomedical Imaging~(ISBI)}. IEEE, 2020, pp. 781--785.

\bibitem{ulyanov2016instance}
D.~Ulyanov et~al.,
\newblock ``{Instance normalization: The missing ingredient for fast stylization},''
\newblock {\em arXiv preprint arXiv:1607.08022}, 2016.

\bibitem{ioffe2015batch}
S.~Ioffe and C.~Szegedy,
\newblock ``{Batch normalization: Accelerating deep network training by reducing internal covariate shift},''
\newblock in {\em International conference on machine learning}. pmlr, 2015, pp. 448--456.

\bibitem{han2020automatic}
S.~Han et~al.,
\newblock ``Automatic cerebellum anatomical parcellation using u-net with locally constrained optimization,''
\newblock {\em Neuroimage}, vol. 218, pp. 116819, 2020.

\bibitem{canny1986computational}
John Canny,
\newblock ``A computational approach to edge detection,''
\newblock {\em IEEE Transactions on pattern analysis and machine intelligence}, , no. 6, pp. 679--698, 1986.

\bibitem{remedios2023self}
S.~W. Remedios et~al.,
\newblock ``Self-supervised super-resolution for anisotropic mr images with and without slice gap,''
\newblock in {\em International Workshop on Simulation and Synthesis in Medical Imaging}. Springer, 2023, pp. 118--128.

\bibitem{zhao2020smore}
C.~Zhao et~al.,
\newblock ``Smore: a self-supervised anti-aliasing and super-resolution algorithm for mri using deep learning,''
\newblock {\em IEEE transactions on medical imaging}, vol. 40, no. 3, pp. 805--817, 2020.

\bibitem{kingma2014adam}
D.P. Kingma and J.~Ba,
\newblock ``{Adam: A method for stochastic optimization},''
\newblock {\em arXiv preprint arXiv:1412.6980}, 2014.

\end{thebibliography}

\end{document}